

\def\squareforqed{\hbox{\rlap{$\sqcap$}$\sqcup$}}
\def\sq{\ifmmode\squareforqed\else{\unskip\nobreak\hfil
\penalty50\hskip1em\null\nobreak\hfil\squareforqed
\parfillskip=0pt\finalhyphendemerits=0\endgraf}\fi}

\def\sun{\hbox{$\odot$}}
\def\la{\mathrel{\mathchoice {\vcenter{\offinterlineskip\halign{\hfil
$\displaystyle##$\hfil\cr<\cr\sim\cr}}}
{\vcenter{\offinterlineskip\halign{\hfil$\textstyle##$\hfil\cr
<\cr\sim\cr}}}
{\vcenter{\offinterlineskip\halign{\hfil$\scriptstyle##$\hfil\cr
<\cr\sim\cr}}}
{\vcenter{\offinterlineskip\halign{\hfil$\scriptscriptstyle##$\hfil\cr
<\cr\sim\cr}}}}}
\def\ga{\mathrel{\mathchoice {\vcenter{\offinterlineskip\halign{\hfil
$\displaystyle##$\hfil\cr>\cr\sim\cr}}}
{\vcenter{\offinterlineskip\halign{\hfil$\textstyle##$\hfil\cr
>\cr\sim\cr}}}
{\vcenter{\offinterlineskip\halign{\hfil$\scriptstyle##$\hfil\cr
>\cr\sim\cr}}}
{\vcenter{\offinterlineskip\halign{\hfil$\scriptscriptstyle##$\hfil\cr
>\cr\sim\cr}}}}}

\def\utw{\smash{\rlap{\lower5pt\hbox{$\sim$}}}}
\def\udtw{\smash{\rlap{\lower6pt\hbox{$\approx$}}}}

\def\diameter{{\ifmmode\mathchoice
{\ooalign{\hfil\hbox{$\displaystyle/$}\hfil\crcr
{\hbox{$\displaystyle\mathchar"20D$}}}}
{\ooalign{\hfil\hbox{$\textstyle/$}\hfil\crcr
{\hbox{$\textstyle\mathchar"20D$}}}}
{\ooalign{\hfil\hbox{$\scriptstyle/$}\hfil\crcr
{\hbox{$\scriptstyle\mathchar"20D$}}}}
{\ooalign{\hfil\hbox{$\scriptscriptstyle/$}\hfil\crcr
{\hbox{$\scriptscriptstyle\mathchar"20D$}}}}
\else{\ooalign{\hfil/\hfil\crcr\mathhexbox20D}}%
\fi}}



\def\bbbc{{\mathchoice {\setbox0=\hbox{$\displaystyle\rm C$}\hbox{\hbox
to0pt{\kern0.4\wd0\vrule height0.9\ht0\hss}\box0}}
{\setbox0=\hbox{$\textstyle\rm C$}\hbox{\hbox
to0pt{\kern0.4\wd0\vrule height0.9\ht0\hss}\box0}}
{\setbox0=\hbox{$\scriptstyle\rm C$}\hbox{\hbox
to0pt{\kern0.4\wd0\vrule height0.9\ht0\hss}\box0}}
{\setbox0=\hbox{$\scriptscriptstyle\rm C$}\hbox{\hbox
to0pt{\kern0.4\wd0\vrule height0.9\ht0\hss}\box0}}}}
\def\bbbq{{\mathchoice {\setbox0=\hbox{$\displaystyle\rm
Q$}\hbox{\raise
0.15\ht0\hbox to0pt{\kern0.4\wd0\vrule height0.8\ht0\hss}\box0}}
{\setbox0=\hbox{$\textstyle\rm Q$}\hbox{\raise
0.15\ht0\hbox to0pt{\kern0.4\wd0\vrule height0.8\ht0\hss}\box0}}
{\setbox0=\hbox{$\scriptstyle\rm Q$}\hbox{\raise
0.15\ht0\hbox to0pt{\kern0.4\wd0\vrule height0.7\ht0\hss}\box0}}
{\setbox0=\hbox{$\scriptscriptstyle\rm Q$}\hbox{\raise
0.15\ht0\hbox to0pt{\kern0.4\wd0\vrule height0.7\ht0\hss}\box0}}}}
\def\bbbt{{\mathchoice {\setbox0=\hbox{$\displaystyle\rm
T$}\hbox{\hbox to0pt{\kern0.3\wd0\vrule height0.9\ht0\hss}\box0}}
{\setbox0=\hbox{$\textstyle\rm T$}\hbox{\hbox
to0pt{\kern0.3\wd0\vrule height0.9\ht0\hss}\box0}}
{\setbox0=\hbox{$\scriptstyle\rm T$}\hbox{\hbox
to0pt{\kern0.3\wd0\vrule height0.9\ht0\hss}\box0}}
{\setbox0=\hbox{$\scriptscriptstyle\rm T$}\hbox{\hbox
to0pt{\kern0.3\wd0\vrule height0.9\ht0\hss}\box0}}}}
\def\bbbs{{\mathchoice
{\setbox0=\hbox{$\displaystyle     \rm S$}\hbox{\raise0.5\ht0\hbox
to0pt{\kern0.35\wd0\vrule height0.45\ht0\hss}\hbox
to0pt{\kern0.55\wd0\vrule height0.5\ht0\hss}\box0}}
{\setbox0=\hbox{$\textstyle        \rm S$}\hbox{\raise0.5\ht0\hbox
to0pt{\kern0.35\wd0\vrule height0.45\ht0\hss}\hbox
to0pt{\kern0.55\wd0\vrule height0.5\ht0\hss}\box0}}
{\setbox0=\hbox{$\scriptstyle      \rm S$}\hbox{\raise0.5\ht0\hbox
to0pt{\kern0.35\wd0\vrule height0.45\ht0\hss}\raise0.05\ht0\hbox
to0pt{\kern0.5\wd0\vrule height0.45\ht0\hss}\box0}}
{\setbox0=\hbox{$\scriptscriptstyle\rm S$}\hbox{\raise0.5\ht0\hbox
to0pt{\kern0.4\wd0\vrule height0.45\ht0\hss}\raise0.05\ht0\hbox
to0pt{\kern0.55\wd0\vrule height0.45\ht0\hss}\box0}}}}
\def\bbbz{{\mathchoice {\hbox{$\sf\textstyle Z\kern-0.4em Z$}}
{\hbox{$\sf\textstyle Z\kern-0.4em Z$}}
{\hbox{$\sf\scriptstyle Z\kern-0.3em Z$}}
{\hbox{$\sf\scriptscriptstyle Z\kern-0.2em Z$}}}}
\def\ts{\thinspace}

\magnification=\magstep0
\font \authfont               = cmr10 scaled\magstep4
\font \fivesans               = cmss10 at 5pt
\font \headfont               = cmbx12 scaled\magstep4
\font \markfont               = cmr10 scaled\magstep1
\font \ninebf                 = cmbx9
\font \ninei                  = cmmi9
\font \nineit                 = cmti9
\font \ninerm                 = cmr9
\font \ninesans               = cmss10 at 9pt
\font \ninesl                 = cmsl9
\font \ninesy                 = cmsy9
\font \ninett                 = cmtt9
\font \sevensans              = cmss10 at 7pt
\font \sixbf                  = cmbx6
\font \sixi                   = cmmi6
\font \sixrm                  = cmr6
\font \sixsans                = cmss10 at 6pt
\font \sixsy                  = cmsy6
\font \smallescriptfont       = cmr5 at 7pt
\font \smallescriptscriptfont = cmr5
\font \smalletextfont         = cmr5 at 10pt
\font \subhfont               = cmr10 scaled\magstep4
\font \tafonts                = cmbx7  scaled\magstep2
\font \tafontss               = cmbx5  scaled\magstep2
\font \tafontt                = cmbx10 scaled\magstep2
\font \tams                   = cmmib10
\font \tamss                  = cmmib10 scaled 700
\font \tamt                   = cmmib10 scaled\magstep2
\font \tass                   = cmsy7  scaled\magstep2
\font \tasss                  = cmsy5  scaled\magstep2
\font \tast                   = cmsy10 scaled\magstep2
\font \tasys                  = cmex10 scaled\magstep1
\font \tasyt                  = cmex10 scaled\magstep2
\font \tbfonts                = cmbx7  scaled\magstep1
\font \tbfontss               = cmbx5  scaled\magstep1
\font \tbfontt                = cmbx10 scaled\magstep1
\font \tbms                   = cmmib10 scaled 833
\font \tbmss                  = cmmib10 scaled 600
\font \tbmt                   = cmmib10 scaled\magstep1
\font \tbss                   = cmsy7  scaled\magstep1
\font \tbsss                  = cmsy5  scaled\magstep1
\font \tbst                   = cmsy10 scaled\magstep1
\font \tenbfne                = cmb10
\font \tensans                = cmss10
\font \tpfonts                = cmbx7  scaled\magstep3
\font \tpfontss               = cmbx5  scaled\magstep3
\font \tpfontt                = cmbx10 scaled\magstep3
\font \tpmt                   = cmmib10 scaled\magstep3
\font \tpss                   = cmsy7  scaled\magstep3
\font \tpsss                  = cmsy5  scaled\magstep3
\font \tpst                   = cmsy10 scaled\magstep3
\font \tpsyt                  = cmex10 scaled\magstep3
\vsize=19.3cm
\hsize=12.2cm
\hfuzz=2pt
\tolerance=500
\abovedisplayskip=3 mm plus6pt minus 4pt
\belowdisplayskip=3 mm plus6pt minus 4pt
\abovedisplayshortskip=0mm plus6pt minus 2pt
\belowdisplayshortskip=2 mm plus4pt minus 4pt
\predisplaypenalty=0
\clubpenalty=10000
\widowpenalty=10000
\frenchspacing
\newdimen\oldparindent\oldparindent=1.5em
\parindent=1.5em
\skewchar\ninei='177 \skewchar\sixi='177
\skewchar\ninesy='60 \skewchar\sixsy='60
\hyphenchar\ninett=-1
\def\newline{\hfil\break}%
\catcode`@=11
\def\folio{\ifnum\pageno<\z@
\uppercase\expandafter{\romannumeral-\pageno}%
\else\number\pageno \fi}
\catcode`@=12 
  \mathchardef\Gamma="0100
  \mathchardef\Delta="0101
  \mathchardef\Theta="0102
  \mathchardef\Lambda="0103
  \mathchardef\Xi="0104
  \mathchardef\Pi="0105
  \mathchardef\Sigma="0106
  \mathchardef\Upsilon="0107
  \mathchardef\Phi="0108
  \mathchardef\Psi="0109
  \mathchardef\Omega="010A
  \mathchardef\bfGamma="0\the\bffam 00
  \mathchardef\bfDelta="0\the\bffam 01
  \mathchardef\bfTheta="0\the\bffam 02
  \mathchardef\bfLambda="0\the\bffam 03
  \mathchardef\bfXi="0\the\bffam 04
  \mathchardef\bfPi="0\the\bffam 05
  \mathchardef\bfSigma="0\the\bffam 06
  \mathchardef\bfUpsilon="0\the\bffam 07
  \mathchardef\bfPhi="0\the\bffam 08
  \mathchardef\bfPsi="0\the\bffam 09
  \mathchardef\bfOmega="0\the\bffam 0A
\def\sun{\hbox{$\odot$}}
\def\la{\mathrel{\mathchoice {\vcenter{\offinterlineskip\halign{\hfil
$\displaystyle##$\hfil\cr<\cr\sim\cr}}}
{\vcenter{\offinterlineskip\halign{\hfil$\textstyle##$\hfil\cr<\cr\sim\cr}}}
{\vcenter{\offinterlineskip\halign{\hfil$\scriptstyle##$\hfil\cr<\cr\sim\cr}}}
{\vcenter{\offinterlineskip\halign{\hfil$\scriptscriptstyle##$\hfil\cr<\cr
\sim\cr}}}}}
\def\ga{\mathrel{\mathchoice {\vcenter{\offinterlineskip\halign{\hfil
$\displaystyle##$\hfil\cr>\cr\sim\cr}}}
{\vcenter{\offinterlineskip\halign{\hfil$\textstyle##$\hfil\cr>\cr\sim\cr}}}
{\vcenter{\offinterlineskip\halign{\hfil$\scriptstyle##$\hfil\cr>\cr\sim\cr}}}
{\vcenter{\offinterlineskip\halign{\hfil$\scriptscriptstyle##$\hfil\cr>\cr
\sim\cr}}}}}
\def\sq{\hbox{\rlap{$\sqcap$}$\sqcup$}}

\def\utw{\smash{\rlap{\lower5pt\hbox{$\sim$}}}}
\def\udtw{\smash{\rlap{\lower6pt\hbox{$\approx$}}}}

\def\diameter{{\ifmmode\mathchoice
{\ooalign{\hfil\hbox{$\displaystyle/$}\hfil\crcr
{\hbox{$\displaystyle\mathchar"20D$}}}}
{\ooalign{\hfil\hbox{$\textstyle/$}\hfil\crcr
{\hbox{$\textstyle\mathchar"20D$}}}}
{\ooalign{\hfil\hbox{$\scriptstyle/$}\hfil\crcr
{\hbox{$\scriptstyle\mathchar"20D$}}}}
{\ooalign{\hfil\hbox{$\scriptscriptstyle/$}\hfil\crcr
{\hbox{$\scriptscriptstyle\mathchar"20D$}}}}
\else{\ooalign{\hfil/\hfil\crcr\mathhexbox20D}}%
\fi}}


\def\bbbc{{\mathchoice {\setbox0=\hbox{$\displaystyle\rm C$}\hbox{\hbox
to0pt{\kern0.4\wd0\vrule height0.9\ht0\hss}\box0}}
{\setbox0=\hbox{$\textstyle\rm C$}\hbox{\hbox
to0pt{\kern0.4\wd0\vrule height0.9\ht0\hss}\box0}}
{\setbox0=\hbox{$\scriptstyle\rm C$}\hbox{\hbox
to0pt{\kern0.4\wd0\vrule height0.9\ht0\hss}\box0}}
{\setbox0=\hbox{$\scriptscriptstyle\rm C$}\hbox{\hbox
to0pt{\kern0.4\wd0\vrule height0.9\ht0\hss}\box0}}}}
\def\bbbe{{\mathchoice {\setbox0=\hbox{\smalletextfont e}\hbox{\raise
0.1\ht0\hbox to0pt{\kern0.4\wd0\vrule width0.3pt height0.7\ht0\hss}\box0}}
{\setbox0=\hbox{\smalletextfont e}\hbox{\raise
0.1\ht0\hbox to0pt{\kern0.4\wd0\vrule width0.3pt height0.7\ht0\hss}\box0}}
{\setbox0=\hbox{\smallescriptfont e}\hbox{\raise
0.1\ht0\hbox to0pt{\kern0.5\wd0\vrule width0.2pt height0.7\ht0\hss}\box0}}
{\setbox0=\hbox{\smallescriptscriptfont e}\hbox{\raise
0.1\ht0\hbox to0pt{\kern0.4\wd0\vrule width0.2pt height0.7\ht0\hss}\box0}}}}
\def\bbbq{{\mathchoice {\setbox0=\hbox{$\displaystyle\rm Q$}\hbox{\raise
0.15\ht0\hbox to0pt{\kern0.4\wd0\vrule height0.8\ht0\hss}\box0}}
{\setbox0=\hbox{$\textstyle\rm Q$}\hbox{\raise
0.15\ht0\hbox to0pt{\kern0.4\wd0\vrule height0.8\ht0\hss}\box0}}
{\setbox0=\hbox{$\scriptstyle\rm Q$}\hbox{\raise
0.15\ht0\hbox to0pt{\kern0.4\wd0\vrule height0.7\ht0\hss}\box0}}
{\setbox0=\hbox{$\scriptscriptstyle\rm Q$}\hbox{\raise
0.15\ht0\hbox to0pt{\kern0.4\wd0\vrule height0.7\ht0\hss}\box0}}}}
\def\bbbt{{\mathchoice {\setbox0=\hbox{$\displaystyle\rm
T$}\hbox{\hbox to0pt{\kern0.3\wd0\vrule height0.9\ht0\hss}\box0}}
{\setbox0=\hbox{$\textstyle\rm T$}\hbox{\hbox
to0pt{\kern0.3\wd0\vrule height0.9\ht0\hss}\box0}}
{\setbox0=\hbox{$\scriptstyle\rm T$}\hbox{\hbox
to0pt{\kern0.3\wd0\vrule height0.9\ht0\hss}\box0}}
{\setbox0=\hbox{$\scriptscriptstyle\rm T$}\hbox{\hbox
to0pt{\kern0.3\wd0\vrule height0.9\ht0\hss}\box0}}}}
\def\bbbs{{\mathchoice
{\setbox0=\hbox{$\displaystyle     \rm S$}\hbox{\raise0.5\ht0\hbox
to0pt{\kern0.35\wd0\vrule height0.45\ht0\hss}\hbox
to0pt{\kern0.55\wd0\vrule height0.5\ht0\hss}\box0}}
{\setbox0=\hbox{$\textstyle        \rm S$}\hbox{\raise0.5\ht0\hbox
to0pt{\kern0.35\wd0\vrule height0.45\ht0\hss}\hbox
to0pt{\kern0.55\wd0\vrule height0.5\ht0\hss}\box0}}
{\setbox0=\hbox{$\scriptstyle      \rm S$}\hbox{\raise0.5\ht0\hbox
to0pt{\kern0.35\wd0\vrule height0.45\ht0\hss}\raise0.05\ht0\hbox
to0pt{\kern0.5\wd0\vrule height0.45\ht0\hss}\box0}}
{\setbox0=\hbox{$\scriptscriptstyle\rm S$}\hbox{\raise0.5\ht0\hbox
to0pt{\kern0.4\wd0\vrule height0.45\ht0\hss}\raise0.05\ht0\hbox
to0pt{\kern0.55\wd0\vrule height0.45\ht0\hss}\box0}}}}
\def\bbbz{{\mathchoice {\hbox{$\sans\textstyle Z\kern-0.4em Z$}}
{\hbox{$\sans\textstyle Z\kern-0.4em Z$}}
{\hbox{$\sans\scriptstyle Z\kern-0.3em Z$}}
{\hbox{$\sans\scriptscriptstyle Z\kern-0.2em Z$}}}}
\def\qed{\ifmmode\sq\else{\unskip\nobreak\hfil
\penalty50\hskip1em\null\nobreak\hfil\sq
\parfillskip=0pt\finalhyphendemerits=0\endgraf}\fi}
\newfam\sansfam
\textfont\sansfam=\tensans\scriptfont\sansfam=\sevensans
\scriptscriptfont\sansfam=\fivesans
\def\sans{\fam\sansfam\tensans}
\def\stackfigbox{\if
Y\FIG\global\setbox\figbox=\vbox{\unvbox\figbox\box1}%
\else\global\setbox\figbox=\vbox{\box1}\global\let\FIG=Y\fi}
\def\placefigure{\dimen0=\ht1\advance\dimen0by\dp1
\advance\dimen0by5\baselineskip
\advance\dimen0by0.33333 cm
\ifdim\dimen0>\vsize\pageinsert\box1\vfill\endinsert
\else
\if Y\FIG\stackfigbox\else
\dimen0=\pagetotal\ifdim\dimen0<\pagegoal
\advance\dimen0by\ht1\advance\dimen0by\dp1\advance\dimen0by1.16666cm
\ifdim\dimen0>\pagegoal\stackfigbox
\else\box1\vskip3.33333 mm\fi
\else\box1\vskip3.33333 mm\fi\fi\fi}
%
\def\begfig#1cm#2\endfig{\par
\setbox1=\vbox{\dimen0=#1true cm\advance\dimen0
by0.83333 cm\kern\dimen0#2}\placefigure}
\def\begdoublefig#1cm #2 #3 \enddoublefig{\begfig#1cm%
\vskip-.8333\baselineskip\line{\vtop{\hsize=0.46\hsize#2}\hfill
\vtop{\hsize=0.46\hsize#3}}\endfig}
\def\begfigsidebottom#1cm#2cm#3\endfigsidebottom{\dimen0=#2true cm
\ifdim\dimen0<0.4\hsize\message{Room for legend to narrow;
begfigsidebottom changed to begfig}\begfig#1cm#3\endfig\else
\par\def\figure##1##2{\vbox{\noindent\petit{\bf
Fig.\ts##1\unskip.\ }\ignorespaces ##2\par}}%
\dimen0=\hsize\advance\dimen0 by-.66666 cm\advance\dimen0 by-#2true cm
\setbox1=\vbox{\hbox{\hbox to\dimen0{\vrule height#1true cm\hrulefill}%
\kern.66666 cm\vbox{\hsize=#2true cm#3}}}\placefigure\fi}
\def\begfigsidetop#1cm#2cm#3\endfigsidetop{\dimen0=#2true cm
\ifdim\dimen0<0.4\hsize\message{Room for legend to narrow; begfigsidetop
changed to begfig}\begfig#1cm#3\endfig\else
\par\def\figure##1##2{\vbox{\noindent\petit{\bf
Fig.\ts##1\unskip.\ }\ignorespaces ##2\par}}%
\dimen0=\hsize\advance\dimen0 by-.66666 cm\advance\dimen0 by-#2true cm
\setbox1=\vbox{\hbox{\hbox to\dimen0{\vrule height#1true cm\hrulefill}%
\kern.66666 cm\vbox to#1true cm{\hsize=#2true cm#3\vfill
}}}\placefigure\fi}
\def\figure#1#2{\vskip0.83333 cm\setbox0=\vbox{\noindent\petit{\bf
Fig.\ts#1\unskip.\ }\ignorespaces #2\smallskip
\count255=0\global\advance\count255by\prevgraf}%
\ifnum\count255>1\box0\else
\centerline{\petit{\bf Fig.\ts#1\unskip.\
}\ignorespaces#2}\smallskip\fi}

\def\begtab#1cm#2\endtab{\par
   \ifvoid\topins\midinsert\medskip\vbox{#2\kern#1true cm}\endinsert
   \else\topinsert\vbox{#2\kern#1true cm}\endinsert\fi}
\def\begpet{\vskip6pt\bgroup\petit}
\def\endpet{\vskip6pt\egroup}
\newcount\frpages
\newcount\frpagegoal
\def\freepage#1{\global\frpagegoal=#1\relax\global\frpages=0\relax
\loop\global\advance\frpages by 1\relax
\message{Doing freepage \the\frpages\space of
\the\frpagegoal}\null\vfill\eject
\ifnum\frpagegoal>\frpages\repeat}
\newdimen\refindent
\def\begrefchapter#1{\titlea{}{\ignorespaces#1}%
\bgroup\petit
\setbox0=\hbox{1000.\enspace}\refindent=\wd0}
\def\ref{\goodbreak
\hangindent\oldparindent\hangafter=1
\noindent\ignorespaces}
\def\refno#1{\goodbreak
\hangindent\refindent\hangafter=1
\noindent\hbox to\refindent{#1\hss}\ignorespaces}
\def\endref{\goodbreak\endpet}
\def\vec#1{{\textfont1=\tams\scriptfont1=\tamss
\textfont0=\tenbf\scriptfont0=\sevenbf
\mathchoice{\hbox{$\displaystyle#1$}}{\hbox{$\textstyle#1$}}
{\hbox{$\scriptstyle#1$}}{\hbox{$\scriptscriptstyle#1$}}}}
\def\petit{\def\rm{\fam0\ninerm}%
\textfont0=\ninerm \scriptfont0=\sixrm \scriptscriptfont0=\fiverm
 \textfont1=\ninei \scriptfont1=\sixi \scriptscriptfont1=\fivei
 \textfont2=\ninesy \scriptfont2=\sixsy \scriptscriptfont2=\fivesy
 \def\it{\fam\itfam\nineit}%
 \textfont\itfam=\nineit
 \def\sl{\fam\slfam\ninesl}%
 \textfont\slfam=\ninesl
 \def\bf{\fam\bffam\ninebf}%
 \textfont\bffam=\ninebf \scriptfont\bffam=\sixbf
 \scriptscriptfont\bffam=\fivebf
 \def\sans{\fam\sansfam\ninesans}%
 \textfont\sansfam=\ninesans \scriptfont\sansfam=\sixsans
 \scriptscriptfont\sansfam=\fivesans
 \def\tt{\fam\ttfam\ninett}%
 \textfont\ttfam=\ninett
 \normalbaselineskip=11pt
 \setbox\strutbox=\hbox{\vrule height7pt depth2pt width0pt}%
 \normalbaselines\rm
\def\vec##1{{\textfont1=\tbms\scriptfont1=\tbmss
\textfont0=\ninebf\scriptfont0=\sixbf
\mathchoice{\hbox{$\displaystyle##1$}}{\hbox{$\textstyle##1$}}
{\hbox{$\scriptstyle##1$}}{\hbox{$\scriptscriptstyle##1$}}}}}
\nopagenumbers
%
\let\header=Y
\let\FIG=N
\newbox\figbox
\output={\if N\header\headline={\hfil}\fi\plainoutput\global\let\header=Y
\if Y\FIG\topinsert\unvbox\figbox\endinsert\global\let\FIG=N\fi}
\let\lasttitle=N
\def\bookauthor#1{\vfill\eject
     \bgroup
     \baselineskip=22pt
     \lineskip=0pt
     \pretolerance=10000
     \authfont
     \rightskip 0pt plus 6em
     \centerpar{#1}\vskip1.66666 cm\egroup}
\def\bookhead#1#2{\bgroup
     \baselineskip=36pt
     \lineskip=0pt
     \pretolerance=10000
     \headfont
     \rightskip 0pt plus 6em
     \centerpar{#1}\vskip0.83333 cm
     \baselineskip=22pt
     \subhfont\centerpar{#2}\vfill
     \parindent=0pt
     \baselineskip=16pt
     \leftskip=1.83333cm
     \markfont Springer-Verlag\newline
     Berlin Heidelberg New York\newline
     London Paris Tokyo Singapore\bigskip\bigskip
     [{\it This is page III of your manuscript and will be reset by
     Springer.}]
     \egroup\let\header=N\eject}
\def\centerpar#1{{\parfillskip=0pt
\rightskip=0pt plus 1fil
\leftskip=0pt plus 1fil
\advance\leftskip by\oldparindent
\advance\rightskip by\oldparindent
\def\newline{\break}%
\noindent\ignorespaces#1\par}}
\def\part#1#2{\vfill\supereject\let\header=N
\centerline{\subhfont#1}%
\vskip75pt
     \bgroup
\textfont0=\tpfontt \scriptfont0=\tpfonts \scriptscriptfont0=\tpfontss
\textfont1=\tpmt \scriptfont1=\tbmt \scriptscriptfont1=\tams
\textfont2=\tpst \scriptfont2=\tpss \scriptscriptfont2=\tpsss
\textfont3=\tpsyt \scriptfont3=\tasys \scriptscriptfont3=\tenex
     \baselineskip=20pt
     \lineskip=0pt
     \pretolerance=10000
     \tpfontt
     \centerpar{#2}
     \vfill\eject\egroup\ignorespaces}
\newtoks\AUTHOR
\newtoks\HEAD
\catcode`\@=\active
\def\author#1{\bgroup
\baselineskip=22pt
\lineskip=0pt
\pretolerance=10000
\markfont
\centerpar{#1}\bigskip\egroup
{\def@##1{}%
\setbox0=\hbox{\petit\kern2.08333 cc\ignorespaces#1\unskip}%
\ifdim\wd0>\hsize
\message{The names of the authors exceed the headline, please use a }%
\message{short form with AUTHORRUNNING}\gdef\leftheadline{%
\hbox to2.08333 cc{\folio\hfil}AUTHORS suppressed due to excessive
length\hfil}%
\global\AUTHOR={AUTHORS were to long}\else
\xdef\leftheadline{\hbox to2.08333
cc{\noexpand\folio\hfil}\ignorespaces#1\hfill}%
\global\AUTHOR={\def@##1{}\ignorespaces#1\unskip}\fi
}\let\INS=E}
\def\address#1{\bgroup
\centerpar{#1}\bigskip\egroup
\catcode`\@=12
\vskip2cm\noindent\ignorespaces}
\let\INS=N%
\def@#1{\if N\INS\unskip\ $^{#1}$\else\if
E\INS\noindent$^{#1}$\let\INS=Y\ignorespaces
\else\par
\noindent$^{#1}$\ignorespaces\fi\fi}%
\catcode`\@=12
\headline={\petit\def\newline{ }\def\fonote#1{}\ifodd\pageno
\rightheadline\else\leftheadline\fi}
\def\rightheadline{\hfil Missing CONTRIBUTION
title\hbox to2.08333 cc{\hfil\folio}}
\def\leftheadline{\hbox to2.08333 cc{\folio\hfil}Missing name(s) of the
author(s)\hfil}
\nopagenumbers
\let\header=Y

\let\lasttitle=N
 \def\contribution#1{\vfill\supereject
 \ifodd\pageno\else\null\vfill\supereject\fi
 \let\header=N\bgroup
 \textfont0=\tafontt \scriptfont0=\tafonts \scriptscriptfont0=\tafontss
 \textfont1=\tamt \scriptfont1=\tams \scriptscriptfont1=\tams
 \textfont2=\tast \scriptfont2=\tass \scriptscriptfont2=\tasss
 \par\baselineskip=16pt
     \lineskip=16pt
     \tafontt
     \raggedright
     \pretolerance=10000
     \noindent
     \centerpar{\ignorespaces#1}%
     \vskip12pt\egroup
     \nobreak
     \parindent=0pt
     \everypar={\global\parindent=1.5em
     \global\let\lasttitle=N\global\everypar={}}%
     \global\let\lasttitle=A%
     \setbox0=\hbox{\petit\def\newline{ }\def\fonote##1{}\kern2.08333
     cc\ignorespaces#1}\ifdim\wd0>\hsize
     \message{Your CONTRIBUTIONtitle exceeds the headline,
please use a short form
with CONTRIBUTIONRUNNING}\gdef\rightheadline{\hfil CONTRIBUTION title
suppressed due to excessive length\hbox to2.08333 cc{\hfil\folio}}%
\global\HEAD={HEAD was to long}\else
\gdef\rightheadline{\hfill\ignorespaces#1\unskip\hbox to2.08333
cc{\hfil\folio}}\global\HEAD={\ignorespaces#1\unskip}\fi
\catcode`\@=\active
     \ignorespaces}
 \def\contributionnext#1{\vfill\supereject
 \let\header=N\bgroup
 \textfont0=\tafontt \scriptfont0=\tafonts \scriptscriptfont0=\tafontss
 \textfont1=\tamt \scriptfont1=\tams \scriptscriptfont1=\tams
 \textfont2=\tast \scriptfont2=\tass \scriptscriptfont2=\tasss
 \par\baselineskip=16pt
     \lineskip=16pt
     \tafontt
     \raggedright
     \pretolerance=10000
     \noindent
     \centerpar{\ignorespaces#1}%
     \vskip12pt\egroup
     \nobreak
     \parindent=0pt
     \everypar={\global\parindent=1.5em
     \global\let\lasttitle=N\global\everypar={}}%
     \global\let\lasttitle=A%
     \setbox0=\hbox{\petit\def\newline{ }\def\fonote##1{}\kern2.08333
     cc\ignorespaces#1}\ifdim\wd0>\hsize
     \message{Your CONTRIBUTIONtitle exceeds the headline,
please use a short form
with CONTRIBUTIONRUNNING}\gdef\rightheadline{\hfil CONTRIBUTION title
suppressed due to excessive length\hbox to2.08333 cc{\hfil\folio}}%
\global\HEAD={HEAD was to long}\else
\gdef\rightheadline{\hfill\ignorespaces#1\unskip\hbox to2.08333
cc{\hfil\folio}}\global\HEAD={\ignorespaces#1\unskip}\fi
\catcode`\@=\active
     \ignorespaces}
\def\motto#1#2{\bgroup\petit\leftskip=5.41666cm\noindent\ignorespaces#1
\if!#2!\else\medskip\noindent\it\ignorespaces#2\fi\bigskip\egroup
\let\lasttitle=M
\parindent=0pt
\everypar={\global\parindent=\oldparindent
\global\let\lasttitle=N\global\everypar={}}%
\global\let\lasttitle=M%
\ignorespaces}
\def\abstract#1{\bgroup\petit\noindent
{\bf Abstract: }\ignorespaces#1\vskip28pt\egroup
\let\lasttitle=N
\parindent=0pt
\everypar={\global\parindent=\oldparindent
\global\let\lasttitle=N\global\everypar={}}%
\ignorespaces}
\def\titlea#1#2{\if N\lasttitle\else\vskip-28pt
     \fi
     \vskip18pt plus 4pt minus4pt
     \bgroup
\textfont0=\tbfontt \scriptfont0=\tbfonts \scriptscriptfont0=\tbfontss
\textfont1=\tbmt \scriptfont1=\tbms \scriptscriptfont1=\tbmss
\textfont2=\tbst \scriptfont2=\tbss \scriptscriptfont2=\tbsss
\textfont3=\tasys \scriptfont3=\tenex \scriptscriptfont3=\tenex
     \baselineskip=16pt
     \lineskip=0pt
     \pretolerance=10000
     \noindent
     \tbfontt
     \rightskip 0pt plus 6em
     \setbox0=\vbox{\vskip23pt\def\fonote##1{}%
     \noindent
     \if!#1!\ignorespaces#2
     \else\setbox0=\hbox{\ignorespaces#1\unskip\ }\hangindent=\wd0
     \hangafter=1\box0\ignorespaces#2\fi
     \vskip18pt}%
     \dimen0=\pagetotal\advance\dimen0 by-\pageshrink
     \ifdim\dimen0<\pagegoal
     \dimen0=\ht0\advance\dimen0 by\dp0\advance\dimen0 by
     3\normalbaselineskip
     \advance\dimen0 by\pagetotal
     \ifdim\dimen0>\pagegoal\eject\fi\fi
     \noindent
     \if!#1!\ignorespaces#2
     \else\setbox0=\hbox{\ignorespaces#1\unskip\ }\hangindent=\wd0
     \hangafter=1\box0\ignorespaces#2\fi
     \vskip18pt plus4pt minus4pt\egroup
     \nobreak
     \parindent=0pt
     \everypar={\global\parindent=\oldparindent
     \global\let\lasttitle=N\global\everypar={}}%
     \global\let\lasttitle=A%
     \ignorespaces}
 \def\titleb#1#2{\if N\lasttitle\else\vskip-28pt
     \fi
     \vskip18pt plus 4pt minus4pt
     \bgroup
\textfont0=\tenbf \scriptfont0=\sevenbf \scriptscriptfont0=\fivebf
\textfont1=\tams \scriptfont1=\tamss \scriptscriptfont1=\tbmss
     \lineskip=0pt
     \pretolerance=10000
     \noindent
     \bf
     \rightskip 0pt plus 6em
     \setbox0=\vbox{\vskip23pt\def\fonote##1{}%
     \noindent
     \if!#1!\ignorespaces#2
     \else\setbox0=\hbox{\ignorespaces#1\unskip\enspace}\hangindent=\wd0
     \hangafter=1\box0\ignorespaces#2\fi
     \vskip10pt}%
     \dimen0=\pagetotal\advance\dimen0 by-\pageshrink
     \ifdim\dimen0<\pagegoal
     \dimen0=\ht0\advance\dimen0 by\dp0\advance\dimen0 by
     3\normalbaselineskip
     \advance\dimen0 by\pagetotal
     \ifdim\dimen0>\pagegoal\eject\fi\fi
     \noindent
     \if!#1!\ignorespaces#2
     \else\setbox0=\hbox{\ignorespaces#1\unskip\enspace}\hangindent=\wd0
     \hangafter=1\box0\ignorespaces#2\fi
     \vskip8pt plus4pt minus4pt\egroup
     \nobreak
     \parindent=0pt
     \everypar={\global\parindent=\oldparindent
     \global\let\lasttitle=N\global\everypar={}}%
     \global\let\lasttitle=B%
     \ignorespaces}
 \def\titlec#1#2{\if N\lasttitle\else\vskip-23pt
     \fi
     \vskip18pt plus 4pt minus4pt
     \bgroup
\textfont0=\tenbfne \scriptfont0=\sevenbf \scriptscriptfont0=\fivebf
\textfont1=\tams \scriptfont1=\tamss \scriptscriptfont1=\tbmss
     \tenbfne
     \lineskip=0pt
     \pretolerance=10000
     \noindent
     \rightskip 0pt plus 6em
     \setbox0=\vbox{\vskip23pt\def\fonote##1{}%
     \noindent
     \if!#1!\ignorespaces#2
     \else\setbox0=\hbox{\ignorespaces#1\unskip\enspace}\hangindent=\wd0
     \hangafter=1\box0\ignorespaces#2\fi
     \vskip6pt}%
     \dimen0=\pagetotal\advance\dimen0 by-\pageshrink
     \ifdim\dimen0<\pagegoal
     \dimen0=\ht0\advance\dimen0 by\dp0\advance\dimen0 by
     2\normalbaselineskip
     \advance\dimen0 by\pagetotal
     \ifdim\dimen0>\pagegoal\eject\fi\fi
     \noindent
     \if!#1!\ignorespaces#2
     \else\setbox0=\hbox{\ignorespaces#1\unskip\enspace}\hangindent=\wd0
     \hangafter=1\box0\ignorespaces#2\fi
     \vskip6pt plus4pt minus4pt\egroup
     \nobreak
     \parindent=0pt
     \everypar={\global\parindent=\oldparindent
     \global\let\lasttitle=N\global\everypar={}}%
     \global\let\lasttitle=C%
     \ignorespaces}
 \def\titled#1{\if N\lasttitle\else\vskip-\baselineskip
     \fi
     \vskip12pt plus 4pt minus 4pt
     \bgroup
\textfont1=\tams \scriptfont1=\tamss \scriptscriptfont1=\tbmss
     \bf
     \noindent
     \ignorespaces#1\ \ignorespaces\egroup
     \ignorespaces}
\let\ts=\thinspace
\def\footnoterule{\kern-3pt\hrule width 1.66666 cm\kern2.6pt}
\newcount\footcount \footcount=0
\def\advftncnt{\advance\footcount by1\global\footcount=\footcount}
\def\fonote#1{\advftncnt$^{\the\footcount}$\begingroup\petit
\parfillskip=0pt plus 1fil
\def\textindent##1{\hangindent0.5\oldparindent\noindent\hbox
to0.5\oldparindent{##1\hss}\ignorespaces}%
\vfootnote{$^{\the\footcount}$}{#1\vskip-9.69pt}\endgroup}
\def\item#1{\par\noindent
\hangindent6.5 mm\hangafter=0
\llap{#1\enspace}\ignorespaces}

\def\titleao#1{\vfill\supereject
\ifodd\pageno\else\null\vfill\supereject\fi
\let\header=N
     \bgroup
\textfont0=\tafontt \scriptfont0=\tafonts \scriptscriptfont0=\tafontss
\textfont1=\tamt \scriptfont1=\tams \scriptscriptfont1=\tamss
\textfont2=\tast \scriptfont2=\tass \scriptscriptfont2=\tasss
\textfont3=\tasyt \scriptfont3=\tasys \scriptscriptfont3=\tenex
     \baselineskip=18pt
     \lineskip=0pt
     \pretolerance=10000
     \tafontt
     \centerpar{#1}%
     \vskip75pt\egroup
     \nobreak
     \parindent=0pt
     \everypar={\global\parindent=\oldparindent
     \global\let\lasttitle=N\global\everypar={}}%
     \global\let\lasttitle=A%
     \ignorespaces}






\def\leaderfill{\kern0.5em\leaders\hbox to 0.5em{\hss.\hss}\hfill\kern
0.5em}
\newdimen\chapindent
\newdimen\sectindent
\newdimen\subsecindent
\newdimen\thousand
\setbox0=\hbox{\bf 10. }\chapindent=\wd0
\setbox0=\hbox{10.10 }\sectindent=\wd0
\setbox0=\hbox{10.10.1 }\subsecindent=\wd0
\setbox0=\hbox{\enspace 100}\thousand=\wd0
\def\contpart#1#2{\medskip\noindent
\vbox{\kern10pt\leftline{\textfont1=\tams
\scriptfont1=\tamss\scriptscriptfont1=\tbmss\bf
\advance\chapindent by\sectindent
\hbox to\chapindent{\ignorespaces#1\hss}\ignorespaces#2}\kern8pt}%
\let\lasttitle=Y\par}
\def\contcontribution#1#2{\if N\lasttitle\bigskip\fi
\let\lasttitle=N\line{{\textfont1=\tams
\scriptfont1=\tamss\scriptscriptfont1=\tbmss\bf#1}%
\if!#2!\hfill\else\leaderfill\hbox to\thousand{\hss#2}\fi}\par}
\def\conttitlea#1#2#3{\line{\hbox to
\chapindent{\strut\bf#1\hss}{\textfont1=\tams
\scriptfont1=\tamss\scriptscriptfont1=\tbmss\bf#2}%
\if!#3!\hfill\else\leaderfill\hbox to\thousand{\hss#3}\fi}\par}
\def\conttitleb#1#2#3{\line{\kern\chapindent\hbox
to\sectindent{\strut#1\hss}{#2}%
\if!#3!\hfill\else\leaderfill\hbox to\thousand{\hss#3}\fi}\par}
\def\conttitlec#1#2#3{\line{\kern\chapindent\kern\sectindent
\hbox to\subsecindent{\strut#1\hss}{#2}%
\if!#3!\hfill\else\leaderfill\hbox to\thousand{\hss#3}\fi}\par}
\long\def\lemma#1#2{\removelastskip\vskip\baselineskip\noindent{\tenbfne
Lemma\if!#1!\else\ #1\fi\ \ }{\it\ignorespaces#2}\vskip\baselineskip}
\long\def\proposition#1#2{\removelastskip\vskip\baselineskip\noindent{\tenbfne
Proposition\if!#1!\else\ #1\fi\ \ }{\it\ignorespaces#2}\vskip\baselineskip}
\long\def\theorem#1#2{\removelastskip\vskip\baselineskip\noindent{\tenbfne
Theorem\if!#1!\else\ #1\fi\ \ }{\it\ignorespaces#2}\vskip\baselineskip}
\long\def\corollary#1#2{\removelastskip\vskip\baselineskip\noindent{\tenbfne
Corollary\if!#1!\else\ #1\fi\ \ }{\it\ignorespaces#2}\vskip\baselineskip}
\long\def\example#1#2{\removelastskip\vskip\baselineskip\noindent{\tenbfne
Example\if!#1!\else\ #1\fi\ \ }\ignorespaces#2\vskip\baselineskip}
\long\def\exercise#1#2{\removelastskip\vskip\baselineskip\noindent{\tenbfne
Exercise\if!#1!\else\ #1\fi\ \ }\ignorespaces#2\vskip\baselineskip}
\long\def\problem#1#2{\removelastskip\vskip\baselineskip\noindent{\tenbfne
Problem\if!#1!\else\ #1\fi\ \ }\ignorespaces#2\vskip\baselineskip}
\long\def\solution#1#2{\removelastskip\vskip\baselineskip\noindent{\tenbfne
Solution\if!#1!\else\ #1\fi\ \ }\ignorespaces#2\vskip\baselineskip}


\long\def\definition#1#2{\removelastskip\vskip\baselineskip\noindent{\tenbfne
Definition\if!#1!\else\
#1\fi\ \ }\ignorespaces#2\vskip\baselineskip}
\def\frame#1{\bigskip\vbox{\hrule\hbox{\vrule\kern5pt
\vbox{\kern5pt\advance\hsize by-10.8pt
\centerline{\vbox{#1}}\kern5pt}\kern5pt\vrule}\hrule}\bigskip}
\def\frameddisplay#1#2{$$\vcenter{\hrule\hbox{\vrule\kern5pt
\vbox{\kern5pt\hbox{$\displaystyle#1$}%
\kern5pt}\kern5pt\vrule}\hrule}\eqno#2$$}
\def\typeset{\petit\noindent This book was processed by the author using
the \TeX\ macro package from Springer-Verlag.\par}
\outer\def\byebye{\bigskip\bigskip\typeset
\footcount=1\ifx\speciali\undefined\else
\loop\smallskip\noindent special character No\number\footcount:
\csname special\romannumeral\footcount\endcsname
\advance\footcount by 1\global\footcount=\footcount
\ifnum\footcount<11\repeat\fi
\gdef\leftheadline{\hbox to2.08333 cc{\folio\hfil}\ignorespaces
\the\AUTHOR\unskip: \the\HEAD\hfill}\vfill\supereject\end}

\def\atsign{@}

%


\contribution{BLACK HOLE, JET, AND DISK: THE UNIVERSAL ENGINE}

\author{Heino Falcke}

\address{Department of Astronomy, University of Maryland, College Park,
MD~20742-2421, USA, email:~hfalcke\atsign{}astro.umd.edu}

\abstract{In this paper I review the results of our ongoing project
to investigate the coupling between accretion disk and radio jet in
galactic nuclei and stellar mass black holes. We find a good
correlation between the {\it UV bump luminosity} and the radio
luminosities of AGN, which improves upon the usual [OIII]/radio
correlations. Taking mass and energy conservation in the jet/disk
system into account we can successfully model the correlation for
radio-loud and radio-weak quasars. We find that jets are comparable in
power to the accretion disk luminosity, and the difference between
radio-loud and radio-weak may correspond to two natural stages of the
relativistic electron distribution -- assuming that radio weak quasars
have jets as well. The distribution of flat- and steep-spectrum
sources is explained by bulk Lorentz factors $\gamma_{\rm
j}\sim5-10$. The absence of radio-loud quasars below a critical
optical luminosity coincides with the FR I/FR II break and could be
explained by a powerdependent, ``closing'' torus. This points towards
a different type of obscuring torus in radio-loud host galaxies which
might be a consequence of past mergers (e.g. by the temporary
formation of a binary black-hole). Interaction of the jet with the
closing torus might in principle also help to make a jet
radio-loud. Turning to stellar-mass black holes we find that galactic
jet sources can be described with the same coupled jet/disk model as
AGN which is suggestive of some kind of universal coupling between jet
and accretion disk around compact objects.  }

\titlea{1}{Introduction}
\titleb{1.1}{The AGN zoo}
Non-stellar activity in galactic nuclei is generally thought to be
produced by a powerful engine located at the dynamical center of the
galaxy. Because of the high luminosity of active galactic nuclei (AGN)
concentrated in a small volume, it has been argued that those engines
are powered by accretion onto a massive black hole.
To the pleasure of observers this activity appears in
many different forms and flavors which has led to a proliferation of
object classes based on specific properties in one or the other wavelength
band. The most important ones are:
\item{a)}{\it Seyfert galaxies} of type 1 (narrow and
broad emission lines) and type 2 (narrow lines only) with luminosities up
to several $10^{44}$ erg/sec, corresponding to accretion rates of $\la 10^{-2}
M_{\sun}/{\rm yr}$
\item{b)}{\it Quasars}, with broad and narrow emission lines and a peak in
their
spectral energy density distribution (SED) in the UV (``UV bump''), the
luminosities are in the range $10^{44..48}$ erg/sec corresponding to
accretion rates of $10^{-2..+2} M_{\sun}/$yr, some quasars have strong
radio emission and are labeled radio-loud, otherwise radio-weak (or
quiet)
\item{c)}{\it Radio galaxies}, with powerful jets, steep-spectrum radio lobes
and
compact, flat-spectrum cores; low-power sources have edge-darkened,
smoke-trail like lobes (FR~I) while high-power sources have well
collimated jets and terminate in hotspots (FR~II), the optical
spectrum of the core usually shows only narrow emission lines
\item{d)}{\it Blazars}, sources completely dominated by variable,
 non-thermal (syn\-chro\-tron) emission from a compact core (BL Lacs),
 if a quasar spectrum is still seen then one has a highly polarized
 quasar (HPQ) or optically violently variable (OVV), those sources
 have the highest probability of showing high-energy ($\ga$GeV)
 emission.

In summary, the different signs of nuclear activity, which are rarely
seen in one object together (except perhaps 3C 273), are: a luminous
thermal bump in the SED, broad and narrow high-excitation emission
lines, a compact radio core, a powerful radio jet and lobes, and
variable high-energy (x-ray to gamma-) emission.

\titleb{1.2}{The unification}
The diversity of objects has provoked the foundation of a
``unification church''\fonote{This religious allusion appears
reasonable considering that both ideas -- despite being basically
correct -- are seemingly not accessible through reasoning but require the
experience of a personal conversion.} (Antonucci 1993 and
refs. therein, see also Antonucci -- this volume) and a sporadic
Counter-Reformation which, however, has not yet rallied its forces
effectively. The ingredients to make the unification work are
relativistic beaming in a jet and obscuration of the central engine by
a molecular torus such that objects appear different if seen from
different aspect angles: a Seyfert 1 galaxy becomes a Seyfert 2 by
obscuration, a radio-loud quasar becomes a FR~II radio galaxy by
obscuration and an HPQ by beaming, likewise a FR~I radio galaxy becomes
a BL Lac by relativistic beaming.

A third ingredient to unify object classes, which I consider very
important but is seldom mentioned explicitly, is the power (or in the
black hole picture the accretion rate) of the nuclear engine. Thus, by
increasing the power, normal galaxies might turn into Seyferts and
then into quasars, provided that the majority of normal galaxies has a
central black hole. As the FR~I/FR~II separation is by morphologhy
{\it and} power, it is likely that by decreasing the accretion rate, a
FR~II radio galaxy turns into an FR~I and then into a quiescent
elliptical. Even though this already gives a pretty and simple
picture, a few fundamental questions remain: a) if high-power (FR~II)
galaxies and quasars, and FR~I and FR~II are connected, how are
low-power (FR~I) and quasars/Seyferts connected, and b) what makes the
difference between a radio-loud and a radio-weak AGN (the \$1,000,000
question)?  Remembering that radio-loud AGN occur only in elliptical
galaxies, the latter question is even more tantalizing. Finally, it
should be realized, that the influence of source evolution is
difficult to assess; shape and size of the obscuring torus may well
depend on the evolutionary stage of the AGN and the host galaxy.

\titleb{1.3}{The universal engine -- a {\it simple} Ansatz}
If the central engine is indeed associated with a black hole, it has
the unpleasant property of being so small that it is almost
inaccessible by observational means and hence open to wild theoretical
speculations.  Currently there is no way to prove or disprove that all
central engines are completely different or absolutely identical and
one is forced to choose a basic Ansatz for the nature of the engine
which allows one to draw further conclusions and test them against
observational data. Strangely, in Astronomy the burden of proof is
usually on those who postulate a simpler solution (like the unified
schemes) while Occam's Razor should force one to start with the
simplest theory until experimentally disproved. Consequently our
Ansatz for the nature of the central engines in AGN should be that
those engines are all very similar and governed by a few parameters
only. Such an engine would be a black hole which accretes matter
within an accretion disk, producing a jet at the black hole/disk
boundary layer flowing out along the rotation axis (alternative
engines are discussed in the contributions by Scheuer, Kundt, and
Sorrell -- this volume). As the escape speed from a black hole is
relativistic, those jets would have to be relativistic as well, as the
jets are produced by the disk, one expects a strong coupling between
jet and disk, and as most of the power in an accretion disk is relased
close to the center, the jet can be very powerful, and finally, as ``a
black hole has no hair'', there are not many parameters that can vary
from one engine to another, and the main parameter of the engine is
expected to be the accretion rate.  This ``simple Ansatz'' also
implies that jets are a natural companion to accretion disks, and both
are necessary and symbiotic features in the accretion process onto the
compact central object (Falcke \& Biermann 1995; hereafter FB95).

\titlea{2.}{Jet-disk coupling}
A coupled jet-disk system has to obey the same conservation laws as
all other physical systems, i.e. energy and mass conservation. We can
express those constraints by specifying that the total jet power
$Q_{\rm jet}$ is a fraction $2q_{\rm j}<1$ of the accretion power
$Q_{\rm disk}=\dot M_{\rm disk}c^2$, the jet mass loss is a fraction
$q_{\rm m}<1$ of the disk accretion rate $\dot M_{\rm disk}$, and the
disk luminosity is a fraction $q_{\rm l}<1$ of $Q_{\rm disk}$
($q_{\rm l}=0.05-0.3$ depending on the spin of the black hole).  The
dimensionless jet power $q_{\rm j}$ and mass loss rate $q_{\rm m}$ are
coupled by the relativistic Bernoulli equation (FB95). For a large
range in parameter space the total jet energy is dominated by its
kinetic energy such that one has $\gamma_{\rm j}q_{\rm m}\simeq q_{\rm
j}$, in case the jet reaches its maximum soundspeed $c/\sqrt{3}$ the
internal energy becomes of equal importance and one has $2\gamma_{\rm
j}q_{\rm m}\simeq q_{\rm j}$ ('maximal jet'). The internal energy is
assumed to be dominated by the magnetic field, turbulence and
relativistic particles. We will constrain the discussion here to the
most efficient type of jet where we have equipartition between the
relativistic particles and the magnetic field and between internal and
kinetic energy -- we will later see that other, less efficient models
(see FB95) would fail.

Knowing the jet energetics, we can describe the longitudinal structure
of the jet by assuming a constant jet velocity (beyond a certain
point) and free expansion according to the sound speed ($\simeq
c/\sqrt{3}$). For such a jet, the equations become very simple, the
magnetic field is given by

$$B_{\rm j}=0.3\,G\;Z_{\rm pc}^{-1}\sqrt{q_{\rm
j/l}L_{46}}$$
and the particle number density is
$$n=11\,{\rm cm}^{-3} L_{46} q_{\rm j/l} Z_{\rm pc}^{-2}$$
(in the jet restframe). Here $Z_{\rm pc}$ is the distance from the
origin in pc, $L_{\rm 46}$ is the disk luminosity in $10^{46}$
erg/sec, $2q_{\rm j/l}=2q_{\rm j}/q_{\rm l}=Q_{\rm jet}/L_{\rm disk}$
is the ratio between jet power (two cones) and disk luminosity which is
of the order 0.3 (FMB95) and $\gamma_{\rm j,5}=\gamma_{\rm j}/5$ ($\beta_{\rm
j}\simeq1$). If one calculates the synchrotron spectrum of such a jet,
one obtains locally a self-absorbed spectrum that peaks at

$$\nu_{\rm ssa}=20\,{\rm GHz}\;{\cal D}{\left(q_{\rm j/l}L_{46}\right)^{2/3}
\over Z_{\rm pc}}\,\left({\gamma_{\rm e,100}
\over\gamma_{\rm j,5} \sin i}\right)^{1/3},$$
integrating over the whole jet yields a flat spectrum with a
monochromatic luminosity of

$$
L_{\nu}={ 1.3\cdot 10^{33}}\,{{\rm erg}\over
{\rm s\, Hz}}\;\left({q_{\rm j/l} L_{46} }\right)^{17/12}
{\cal D}^{13/6}\sin i^{1/6} \gamma_{\rm
e,100}^{5/6} \gamma_{\rm j,5}^{11/6},
$$
where $\gamma_{\rm e,100}$ is the minimum {\it electron} Lorentz
factor divided by 100, and ${\cal D}$ the {\it bulk} jet Doppler
factor. At a redshift of 0.5 this luminosity corresponds to an
unboosted flux of $\sim100$ mJy. The brightness temperature of the jet
is

$${T}_{\rm b}=1.2\cdot 10^{11}\, {\rm K}\; {\cal D}^{4/5}{\left({
{\gamma_{\rm e,100}}^2 q_{\rm j/l} L_{46} \over
\gamma_{\rm j,5}^2 \beta_{\rm j}}\right)^{1/12}\sin i^{5/6}}
$$
which is almost independent of all parameters except the Doppler
factor. An important factor that governs the synchrotron emissivity is
of course the electron distribution, for which we have assumed a
powerlaw distribution with index $p=2$ and a ratio 100 between maximum
and minimum electron Lorentzfactor. As we are discussing here the most
efficient jet model we also assume that all electrons are accelerated
(i.e. $x_{\rm e}=1$ in FB95), hence the only remaining parameter is
the minimum Lorentzfactor of the electron distribution $\gamma_{\rm
e,100}$ determining the total electron energy content. In order to
reach the magnetic field equipartition value, which is close to the
kinetic jet power governed by the protons, we have to require
$\gamma_{e,100}\sim1$. It cannot be higher because otherwise the power
in electrons would exceed the total jet power, and it cannot be much
lower because we would not reach equipartition. Such a high,
low-energy cut-off in the electron energy distribution was suggested
already by Wardle (1977) and Celotti \& Fabian (1993) for other
reasons.
\medskip

 If radio-interferometric techniques were not yet developed today,
and we would have been asked to predict what kind of jet sources we
would expect to see, we would have needed only very few simple
considerations:
\item{a)} {\it `total equipartition' everywhere}, i.e. equipartition between
the luminosity radiated by the disk and expelled by a jet wind,
equipartition between internal energy and kinetic energy, and
equipartition between relativistic particles and magnetic field
\item{b)} {\it relativistic speed}, because,  if the jet is  produced
close to the black hole, relativistic escape speeds are required,
\item{c)} {\it disk luminosity} (UV-bump), which is a measurable quantity

Thus, using $L_{\rm disk}\sim10^{46}$ erg/sec and $\gamma_{\rm j}\sim
5$, we could have predicted pc-scale radio cores at cm-wavelengths,
with brightness temperatures of $10^{11}$ K and fluxes of 100 mJy and
more.  But of course, nobody would have believed us as those
assumptions are obviously too simplified\dots

\titlea{3.}{UV/radio correlation}
\titleb{3.1}{Estimating the disk luminosity}
Now, we will have to validate some of our assumptions and
test the jet-disk coupling derived above. For this we have to estimate
the disk luminosity of quasars as precisely as possible and compare it
to their radio cores. The best studied quasars sample so far is the PG
quasar sample (Schmidt \& Green 1983). For most sources in this sample
Sun \& Malkan (1989), using optical and IUE data, fitted the UV bump
with accretion-disk models and a few more were available in the
archive (Falcke, Malkan, Biermann 1995, hereafter FMB95). There are
also excellent photometric (Neugebauer et al. 1987) and spectroscopic
data (Boroson \& Green 1992) available, but unlike the broadband
UV-bump fits, emission lines and continuum colors do not give a direct
estimate for the bolometric UV luminosity ($L_{\rm disk}$) and we need
to calibrate those values to the UV bump luminosity using the sources
which have a complete set of data available, yielding

$$\lg (L_{\rm disk}/{\rm erg\;s}^{-1})=2.85+\lg (L_{\rm [OIII]}/{\rm
erg\;s}^{-1}),$$
$$\lg (L_{\rm disk}/{\rm erg\;s}^{-1})=2.1+\lg (L_{\rm H{\beta}}/{\rm
erg\;s}^{-1}),$$
$$\lg (L_{\rm disk}/{\rm erg\;s}^{-1})=-0.4 M_{\rm b}+35.90.$$

With those correlations one should be able to estimate the ``disk
luminosity'' for almost any quasar. If several indicators are
available, we can combine them (assigning appropriate weights) to get
the final estimate, this then gives a fairly reliable estimate of
$L_{\rm disk}$ and reduces the scatter in the correlations
considerably as it also reduces the effects of the orientation
dependence of some lines (e.g. [OIII]).  In the next step, we can
compare those disk luminosities with VLA radio cores (Kellermann et
al. 1989, Miller et al. 1993) and total radio emission. In addition to
the optically selected sample in FMB95 I have now also included
quasars from the southern 2 Jansky sample (Morganti et al. 1994,
Tadhunter et al. 1994) which are predominantly flat-spectrum quasars,
and steep-spectrum, lobe dominated quasars from Bridle et al. (1994),
Akujor et al. (1995), and Reid et al. (1995) which had emission lines
readily available (Steiner 1981, Jackson \& Browne 1991, Wills et
al. 1993). Thus, the number of radio-loud quasars is increased
considerably -- the results are shown in Figures 1\&2.

\begfig 8.2cm
\figure{1}{Total radio luminosity vs. disk (UV-bump) luminosity for
quasars (including a complete optical and a radio-selected
sample). The shaded circles are core-dominated, flat-spectrum sources,
open circles are steep-spectrum (FR II type) sources, circles labeled
'c' are CSS sour\-ces and filled points are radio-weak (diffuse or
unresolved) sources. Only un\-dis\-turbed radio-loud FR II type and
radio-weak sources show a tight correlation. Flat-spectrum sources are
boosted to the upper end of the distribution except 6
radio-intermediate quasars which might be boosted radio-weak quasars
(see text). The total emission of CSS does not show a tight
correlation with disk luminosity. The solid line is the
(oversimplified) model for the lobes from FB95 (see also FMB95).}
\endfig

\titleb{3.2}{Different types of radio sources}
For those kinds of optical/radio correlation it is very important what
kind of radio source one is talking about. Here I distinguish between
four cases represented by different symbols: a) radio-weak quasars
with weak, diffuse or unresolved radio emission, b) core dominated,
flat-spectrum sources, c) FR\'II type steep spectrum sources, and d)
compact steep spectrum (CSS) or irregular radio sources. It is quite
obvious that the radio/optical correlation may be very different for
all these sources -- at least in total flux. CSS
and irregular sources usually have jets which are strongly interacting
with a dense environment inside the galaxy, and their radio output is
expected to be strongly modified by this interaction. Flat spectrum
sources are usually dominated by their relativistically boosted jet
with the inclination being a very sensitive parameter and cannot be
compared with the steep-spectrum emission of lobe-dominated sources,
and finally steep-spectrum radio-loud and regular radio-weak sources
have a completely different radio morphology and hence must also be
treated separately.

This is highlighted in Fig. 1, where one can see that radio-loud and
radio-weak sources are clearly separated. The undisturbed,
steep-spectrum FR II sources do show a relatively tight correlation,
the CSS and irregular quasars scatter around and do not show any
correlation, and the core-dominated sources are mainly located at the
upper end of the radio distribution consistent with being
relativistically boosted -- with the exception of a few
radio-intermediate quasars (RIQ) located in the gap between radio-loud
and radio-weak quasars.

\titleb{3.3}{Boosted radio-weak quasars?}
The RIQ in the gap are all optically selected PG quasars, and their
radio fluxes are typically a few ten up to a few hundred mJy. From
their $R$-ratio between radio and optical flux, they are neither
clearly radio-loud nor clearly radio-weak, but all are unresolved with
the VLA. Using the Effelsberg 100m telescope, we measured the fluxes
of those sources at 11 and 2.8 cm to look for variability and spectral
slopes (Falcke, Sherwood, Patnaik 1996, in prep.). Even though the
data is not yet fully evaluated it is quite obvious that 6 of those
sources have flat-spectrum cores and at least 5 are variable. In order
to distinguish them from the 2 probable CSS sources which can also be
found in the radio-intermediate 'gap', we will label them
flat-spectrum radio-intermediate quasars (FIQ).

Some of the FIQ were known as variable sources before, e.g. III Zw 2
which varies between 40 mJy and 1 Jy at 5 GHz and has a brightness
temperature probably well in excess of $10^{11}$K (Ter\"asanta \&
Valtaoja 1994)-- ususally a sign of relativistic boosting. The
presence of a variable flat-spectrum radio core (without or with weak
extended emission) alone is usually already regarded as a good sign
for relativistic boosting. In FMB95 (with the knowledge of only the 3
FIQs in the $z<0.5$ sample), we have argued that, if the variability
and the core prominence is due to relativistic boosting, it is
unlikely that those quasars are boosted radio-loud cores: at a given
optical luminosity their allegedly boosted radio-cores are much weaker
than the usual distribution of flat-spectrum radio quasars and as
bright or even weaker than the cores of (unboosted) lobe-dominated
quasars. Their total flux is also much lower than the total flux of
radio-loud quasars, demonstrating that -- like radio weak quasars --
they lack anything similar to the radio lobes expected for radio-loud
quasar. Hence, the only parent population they could have been boosted
from are the radio-weak quasars. As shown in FMB89 the relative number
of FIQs and their offset from the parent population is consistent with
moderate Lorentzfactors of 3-5. An interesting test for the radio-weak
blazar hypothesis for the FIQ will be VLBI observations and an
investigation of their host galaxies -- one would expect to find at
least a few of them in spiral host galaxies, as opposed to radio-loud
host galaxies which are exclusively in ellipticals.

\titleb{3.4}{Modelling the UV/radio distribution}
The fact that we find such good correlations between the UV-bump and
the radio luminosities also tells us a lot. Especially for the
radio-weak quasars it shows clearly that the nuclear and the {\it
extended} radio emission is AGN related.  It would be very difficult
to find an argument that the extended emission is produced by
starbursts and explain the tight UV/radio correlation unless one is
willing to postulate that the UV itself is produced by a starburst (as
Terlevich et al.~1992). The total radio emission of the undisturbed,
lobe-dominated radio-loud quasars, which is clearly jet-related,
scales with the UV-bump as well, which indeed suggests a direct link
between the radio-jet producing mechanism and the UV source.

We can now use the jet-disk model derived in Sec. 2 and compare it to
the UV/radio correlation for the cores. Because we have carefully
calibrated the line emission and the continuum fluxes and scaled them
to the UV-bump luminosity, we are able to apply an actual physical
model with absolute numbers to the distribution and hence can apply
the mass and energy conservation laws to it.

\begfig 8.2cm
\figure{2}{The same as Fig. 1 (CSS are not explicitly marked) but now
the radio core flux is plotted. The shaded bands represent the
radio-loud and radio-weak jet model where the width is determined by
relativistic boosting. The dashed line represents sources at the
boosting cone (inclination $1/\gamma$) and the solid line represents
$0^\circ$ inclination -- corresponding to the maximum possible
flux. The position of flat-spectrum and steep-spectrum sources and the
ra\-dio-loud/radio-weak separation can be naturally accounted for with
the coupled jet-disk model plus boosting.}
\endfig

To simplify the discussion, we will use only the most efficient model,
where the internal energy is comparable to the kinetic energy and
dominated by the magnetic field and relativistic particles. If we
would start with a normal plasma jet, where the number of particles is
limited by the mass conservation, and assume that all electrons are
accelerated from the thermal pool ($\gamma_{\rm e}\sim1$) into a
powerlaw distribution, we find that we can well explain the radio
luminosity of the radio-weak quasars (Fig. 2, lower band), however,
fail to explain the radio-loud quasars by a large margin if we demand
$2q_{\rm j/l}<1$. The reason for this is that in such a model the total
energy of the electrons is still just a small fraction of the total
energy dominated by the kinetic energy of the ions (``protonic
model''), and most of the electrons are found at low energies where
they do not contribute to the radio flux. To bring the electrons in
equipartition, one either has to create additional pairs (100 times
more then electrons) or inject them at a high energy where
$\gamma_{\rm e}$ is of the order 100 (``electronic model''). Such
models are the only ones, which are capable of explaining the UV/radio
correlation for radio-loud quasar cores and they do require the `total
equipartition' mentioned in Section 2.

Interestingly, related energetical arguments earlier have led the editor of
this book to speculate that high electron Lorentz factors must be
present in radio jets (Kundt \& Gopal-Krishna 1980). He interpreted
this as evidence for ultrarelativistic bulk Lorentz factors, but as
demonstrated in Fig. 2 the spread of radio core luminosities in jets
(due to the anisotropy of relativistic boosting) is too narrow for
such high bulk Lorentz factors and therefore those Lorentz factors
must indicate internal, random motions of the electrons.

In order to reproduce the whole distribution of the radio cores, with
the two equipartition models (electronic and protonic), we have to
specify only two parameters: the jet-disk ratio $q_{\rm j/l}$ which we
assume to be constant, and the proper velocity of jet for which we make
the powerlaw Ansatz $\gamma_{\rm j}\beta_{\rm
j}=6((2/6)^{1/0.15}+{L_{46}})^{0.15}$. This allows a moderate increase
of the jet velocity with power, where the typical quasar Lorentzfactor
is $\gamma_{\rm j}\sim6$ at $L_{46}=1$, but never becomes
subrelativistic, i.e. smaller than $\gamma_{\rm j}\sim2$. In Fig. 2
the two equipartition models are shown, depicting the regions within
the boosting cone (dashed and solid line) and the unboosted population
(shaded band); the jet/disk parameter used here is $2q_{\rm j/l}=0.3$
(two-sided jet).

Many conclusions can be drawn from this simple kind of analysis:
\item{a)} The Lorentzfactors have moderate values between 5-10, and there
is no evidence for {\it stationary bulk} Lorentzfactors far in excess
of 10 in quasars. Those sources -- if seen face on -- should have a
radio/UV ratio much higher than seen in any of the sources in our
sample.
\item{b)} Radio-loud sources are utmost efficient jets, and the differences
between radio-loud and radio-weak sources are remarkably close to the
difference between two natural stages of the electron distribution:
one starting at thermal energies, the other shifted up in
energy-space until equipartition is reached.
\item{c)} The radio jets have powers comparable to the disk luminosity, hence
they must be produced in the very inner parts of the disk close to the
black hole where the bulk of the gravitational energy is released.
\item{d)}The magnetic flux in the jet is much higher than the maximum
possible radial magnetic flux in an accretion disk and hence the
magnetic field for the jet must be produced locally at the footpoint
and because of the high efficiency must be related somehow to the
dissipation process in the disk (FB95).

\titlea{4.}{Unified unification}
\titleb{4.1}{The void of FR I quasars}
There is another very interesting observation to be made in
Figure~1. While the radio-quiet quasars spread out over the whole
luminosity interval from $10^{44}-10^{48}$ erg/sec, radio-loud quasars
are predominantly found in the range $L>10^{46}$ erg/sec, and none is
below $L\sim2\cdot10^{45}$ erg/sec. One may argue that at the lower
luminosities, the larger elliptical host galaxies become visible and
therefore the sources are not classified as quasars, however, this
falls short of explaining the void over 2 orders of magnitude. The
alternative explanation is that in fact below a critical power,
radio-loud quasars lose their typical quasar characteristics, i.e.
the broad (and narrow) emission lines and the UV bump. A hint why this
may be so comes from the radio morphological data we have for the PG
quasars: all radio-loud sources are either of FR II type or compact,
none has a typical FR I structure, and indeed it is part of the
radio-astronomers folklore that FR I radio sources never show up as
quasars.  {\it Consequently, we can identify the void of radio-loud
quasars below a certain optical luminosity with the FR II (high-power)
to FR I (low-power) transition.} This link between transition of radio
morphology and disappearence of optical emission is difficult to
understand, especially as the emission line properties of radio-weak
quasars -- which are almost identical to those of radio-loud quasars
-- do not show any change at this critical power. Thus just a change
(or disappearance) of the accretion disk with decreasing power seems
unlikely. A change of the engine would also violate the simplicity of
our Ansatz and therefore is not the preferred option here.

\titleb{4.2}{The closing torus}
The question now is whether we can explain the behaviour of the
radio-loud sources qualitatively without having to postulate different
central engines. A minor modification to the unified scheme may
indeed do this job: if the opening angle of the obscuring torus is not
constant but power-dependent, the torus could approach the jet opening
angle at low powers. In this case the central engine would be obscured
for almost all aspect angles, and the interaction between the jet and
the torus could start the disruption of the low-power jet and initiate
its morphological transition (Falcke, Gopal-Krishna, \& Biermann 1995).

\titleb{4.3}{Observational consequences}
Consequently we would not expect to see broad emission lines from a FR
I type radio source as the broad-line region is expected to be inside
the 1-100 pc scale torus and be completely obscured. If the opening
angle is smaller than the boosting cone, even for boosted FR Is
(i.e. BL Lacs) most of the emission line region would be obscured. In
fact, one would have to wonder if broad emission line clouds could
survive at all in the narrow funnel of the torus if it extends down to
the smallest scales.  The narrow lines would also be strongly
suppressed as the escaping ionizing continuum that produces the NLR
region itself is suppresssed. The same is true for broad lines in polarized
(scattered) light: as there is not much optical light escaping from
the nucleus there will also not be much light that could be scattered.

The best wavelength regime to test the ``closing torus'' scenario
would therefore be the IR where most of the energy of the central
source should be re-emitted, predominantly at 10-20$\mu$m. The FR I
should therefore have an IR output comparable to FRII radio galaxies
and quasars scaled to the same engine power. As already in FR II
galaxies and quasars, more than half of the energy is absorbed, the
{\it relative} increase in the IR luminosity for FR I, where almost
everything is absorbed, is less than a factor two and difficult to
detect. However, because of the different shapes of the tori, the IR
spectrum itself (e.g. $10\mu$m silicate feature) might be different
(see Pier \& Krolik 1992). Also NIR spectroscopy might reveal the
presence of a quasar engine in FR I radio galaxies; a first pilot
study is currently on the way.

Another observational effect concerns the ratio between quasars and
radio galaxies in low-radio-frequency selected (orientation
independent) samples. This ratio should not be constant but depend on
the power of the central engine as it reflects the width of the torus
opening. Such a powerdependence is indeed observed (and often used as
an argument against the quasar/galaxy unification; Lawrence 1991;
Singal 1993).

\titleb{4.4}{Making the jet radio-loud}
The fact that the jet may interact with the torus opens the field for
many interesting speculations and future studies. If we imagine a
powerful, magnetized, relativistic jet scraping along the inner parts
of a dense torus (or a cloud therein, or a cloud blown off its surface)
we may anticipate the formation of a violent shear layer between jet
and the external medium. The interaction might induce highly oblique
shocks where particles are accelerated and thus would make the jet
radio loud. If the shear-layer is very thin, magnetized and dense,
collisions of the ions, carried by the jet at relativisitic speeds,
with the external matter would lead to hadronic cascades and
consequently to the production of gamma rays and to the injection of
pairs at $\ga35$ MeV -- this could also make the jet radio-loud.

In this context it is interesting to note that AGN jets cannot start
as radio-loud jets. The synchrotron losses and the inverse Compton
losses of the relativistic particles with the UV photons from the disk
would be catastrophic and lead to a complete dissipation of the
relativistic electrons with high Lorentz factors in its inner parts. A
simple extrapolation of the radio emission to the black hole scale
would also predict a radio luminosity in excess of the jet power.
The typical scale where the losses become less severe and the
electron injection can happen is $z\ga10^{16-17}$ cm (FB95). This is
close to the scale often quoted for the production of gamma rays and
the scale for the BLR. One starts to wonder whether gamma-ray
emission, electron injection and jet torus or jet/BLR interaction have
something to do with each other.

\titleb{4.5}{Torus and host galaxy}
In the whole discussion we have ignored the nature of the torus and
its origin and why it should be powerdependent. None of these
questions can be readily answered. The torus may be just a very
thick accretion disk with a steep funnel, or it may be composed of
molecular clouds in turbulent motion. The powerdependence of its
opening might be due to heating and depletion of its inner parts by
the central engine or due to the jet itself that drills through a more
or less sphercial dust distribution and carries matter outwards,
thereby opening the ``torus''.

In any case a very high (gas, magnetic, or turbulent) pressure is
needed to maintain the thickness of the torus.  Besides the electron
injection discussed above, interaction of the jet with the torus
itself, or with winds or clouds produced by it, may therefore also
have a confining effect at the inner torus scale (sub-pc to pc). A jet
without such a closing torus and without the jet/torus interaction
would neither be well collimated nor have the efficient injection of
electrons/pairs to make it radio-loud. {\it Hence it is in principle
possible to attribute the radio-loud/radio-quiet dichotomy to
different environments at the pc scale rather than to differences in
the engine itself.}

The advantage of this concept is that it is easier to relate the pc
scale environment rather than the engine properties to the host
galaxy. It was proposed that the merging of galaxies may lead to a
spin-up of the central black hole by black hole merger during the
creation of an elliptical galaxy (Wilson \& Colbert 1995). However, it
is by no means clear why spiral galaxies should not have a rotating
black hole ab initio. Nevertheless, the idea of mergers being a
necessary prerequisite for the production of a radio-loud jet is quite
tempting and supported by observations (e.g. Heckman et al. 1986). And
indeed it seems unavoidable that such a merger sooner or later leads
to the formation of a single black hole. There is, however, a critical
separation between the two merging holes -- again at the pc scale --
where neither gravitational friction nor gravitational radiation is
very efficient (Begelman, Blandford, Rees 1980), and the binary may
stay there for quite a while. So, merging may lead to black hole
coalescence, but it definitively will also change the pc-scale
structures of the stars and the dust in the central bulge in a way
which will be very different from those in spiral galaxies!

\titlea{5.}{Starved and stellar-mass black holes}
\titleb{5.1}{Sgr A* and its siblings}
An interesting consequence of our approach is that it is to first
order independent of the scale. If jets and disks are symbiotic
features, this may apply to almost any kind of accretion disk with a
compact central mass. The equations in Sec. 2 do not depend on the
mass of the central object but mainly on the mass accretion rate (i.e.
the disk luminosity) and consequently we can use the same scheme for
stellar-mass black holes and for the starved black holes in inactive
galactic nuclei such as in our own Galaxy. In fact, the jet/disk
symbiosis was initially developed to explain the Galactic Center
source Sgr A* (Falcke et al. 1993a\&b; Falcke 1996a\&b) and the
jet/disk model still remains a viable explanation for this compact
source. Also the radio core of other weakly active galaxies like M81
and M31 seem to follow the same rules (see Falcke 1994, Figure 8.1).

\titleb{5.2}{Galactic jet sources}
The sources which received most attention recently are the
galactic jet sources associated with either neutron stars or black hole
candidates. Two of those sources show apparent superluminal motion
indicating relativistic speeds, and they have properties similar to
extragalactic jets (Mirabel \& Rodriguez 1994; Hjellming \& Rupen
1995). Besides those sources, we also know a few other sources in the
Galaxy (1E1740-2942, SS433, GRS 1758-258) which show clear jets but
without superluminal motion yet detected, and a few x-ray
binaries do show flat spectrum radio cores which might be related to a
jet. To compare the radio cores of all these sources with their disk
luminosity, one has to use x-ray data because the disk spectrum of low-mass
black holes is shifted to higher energies. In Fig. 3 the galactic jet
sources are shown in a diagram similar to Fig. 2 but extending now
down to very low luminosities (Falcke \& Biermann 1996). As one can
see, no change of the basic parameters is necessary to roughly
predict the range of radio fluxes expected for these sources at a
luminosity which is 6-10 orders of magnitude lower than in the
supermassive AGN black holes. There is even a hint for a dichotomy
between radio-loud and radio-weak sources among the galactic jets, but
the statistics are not yet good enough. The fact that those equations
and the formulation of the symbiosis principle that implied the
presence and the luminosity of the galactic superluminal jet sources
(FB95; Falcke 1994) were suggested {\it before} the discovery
of those sources, demonstrates the predictive power of this
principle --- and there is yet a lot of parameter space in Fig. 3 to be
filled.
\begfig 8.2cm
\figure{3}{The same model and data as in Fig. 2 for quasars but now extended
to lower powers where galactic jet sources are found. Here the disk
luminosity corresponds is interpreted as the x-ray luminosity. Big
stars represent confirmed jet sources, while small stars represent
x-ray binaries. The big black dots are Sgr A* and M31*.}
\endfig

\titlea{6.}{Summary}
This work could be summarized by stating that at present it is {\it
not possible} to show observationally that the central engines in AGN
are essentially different (i.e. on a scale of 10-100 $R_{\rm
g}$). Postulating that jets and disks around compact objects are
symbiotic and universal features is sufficient to account for most of
the observed effects and allows several interesting conclusions:
\item{$\bullet$} The jet/disk coupling explains the UV/radio
correlation for quasars and the x-ray/radio flux of stellar-mass black
holes.
\item{$\bullet$} Radio-loud jets are utmost efficient and
have total powers comparable to the disk luminosities.
\item{$\bullet$} The distribution of flat- and steep-spectrum quasars within
the
UV/radio correlation reflects relativistic boosting with $\gamma_{\rm
j}\sim5$, and a population of flatspectrum radio-intermediate quasars
(FIQ) suggests the presence of relativistic jets in radio-weak quasars.
\item{$\bullet$} The pc-scale environment (``the torus'') may change the jet
properties
drastically. For example, a closing torus in radio-loud galaxies may
explain the transition from FR II jets to FR I jets and the weakness
of emission lines in FR I and BL Lacs. The jet/torus interaction may
in principle also help to make a jet radio-loud.
\item{$\bullet$} There is no fundamental difference in the parameters between
jets from
stellar-mass und supermassive black holes.

The fact that stellar-mass black holes and AGN can be described with
the same simple jet/disk model (Fig. 3) suggests that there is a
universal correlation between radio emision and disk luminosity that
spans the whole luminosity range from a few hundred solar luminosities
up to several $10^{14}L_{\sun}$, and hence there may be quite a few other
sources (e.g. Seyfert galaxies and nearby galactic nuclei) that follow
the same trend but have not yet been discussed in this respect.

\bigskip\noindent

\begrefchapter{References}
\ref Akuyor C.E., L\"udke E., Browne I. et al. 1994, A\&AS 105, 247
\ref Bridle A.H., Hough D.H., Lonsdale C.J., Burns J.O., Laing R.A. 1994, AJ
108, 766
\ref Antonucci R. 1993, ARAA 31, 473
\ref Antonucci R. 1996 -- this volume
\ref Begelman M.C., Blandford R.D., Rees M.J. 1980, Nat 287, 307
\ref Boroson T.A., Green R.F. 1992, ApJS 80, 109
\ref Celotti A., Fabian A. 1993, MNRAS 264, 228
\ref Falcke H. 1994, PhD thesis, RFW Universit\"at Bonn
\ref Falcke H. 1996a, to appear in:  ``Unsolved Problems of the
Milky Way'', IAU Symp. 169, L. Blitz \& P.J. Teuben (eds.), Kluwer, Dordrecht,
p. 163
\ref Falcke H. 1996b -- this volume
\ref {Falcke, H., Biermann, P. L. 1995, A\&A 293, 665 (FB95)}
\ref {Falcke, H., Biermann, P. L. 1996, A\&A in press}
\ref {Falcke, H., Biermann, P. L., Duschl, W. J., Mezger, P. G. 1993a, A\&A
270, 102}
\ref {Falcke, H., Mannheim, K., Biermann, P. L. 1993b,A\&A 278, L1}
\ref {Falcke, H., Malkan, M., Biermann, P.L. 1995a, A\&A 298, 375 (FMB95) }
\ref {Falcke, H., Gopal-Krishna, Biermann, P.L. 1995b, A\&A 298, 395}
\ref Heckman T.M., Smith E.P., Baum S.A. et al. 1986, ApJ 311, 526
\ref Hjellming R. M., Rupen M.P. 1995, Nat 375, 464
\ref Jackson N., Browne I.W.A. 1991, MNRAS 250, 414
\ref Kellermann K.I., Sramek R., Schmidt M., Shaffer D.B., Green R. 1989, AJ
98, 1195
\ref Kundt W., Gopal-Krishna 1980, Nat 288, 149
\ref Lawrence A. 1991, MNRAS 252, 586
\ref Miller P., Rawlings S., Saunders R. 1993a, MNRAS 263, 425 (MRS)
\ref {Mirabel I.F., Rodriguez 1994, Nat 371, 46 }
\ref Neugebauer G., Green R.F., Matthews K. et al. 1987, ApJS 63, 615
\ref Pier E., Krolik J.H. 1992, ApJ 401, 99
\ref Reid A., Shone D.L., Akujor C.E. et al. 1995, A\&AS 110, 213
\ref Schmidt M., Green R. 1983, ApJ 269, 352
\ref Steiner J.E. 1981, ApJ 250, 469
\ref Sun W.H., Malkan M.A. 1989, ApJ 346, 68 (SM89)
\ref Singal A.K. 1993, MNRAS 262, L27
\ref Tadhunter C.N., Morganti R., di Serego Alighieri S., Fosbury R.A.E.,
     Danziger I.J. 1993, MNRAS 263, 999
\ref Ter\"asanta, H., Valtaoja, E. 1994, A\&A 283, 51
\ref Terlevich R., Tenoria-Tagle G., Franco J., Melnick J. 1992, MNRAS 255, 713
\ref Wardle J.F.C. 1977, Nat 269, 563
\ref Wills B.J., Netzer H., Brotherton M.S., et al. 1993, ApJ 410, 534
\ref Wilson A.S. \& Colbert E.  1995, ApJ 438, 62
\endref

\byebye
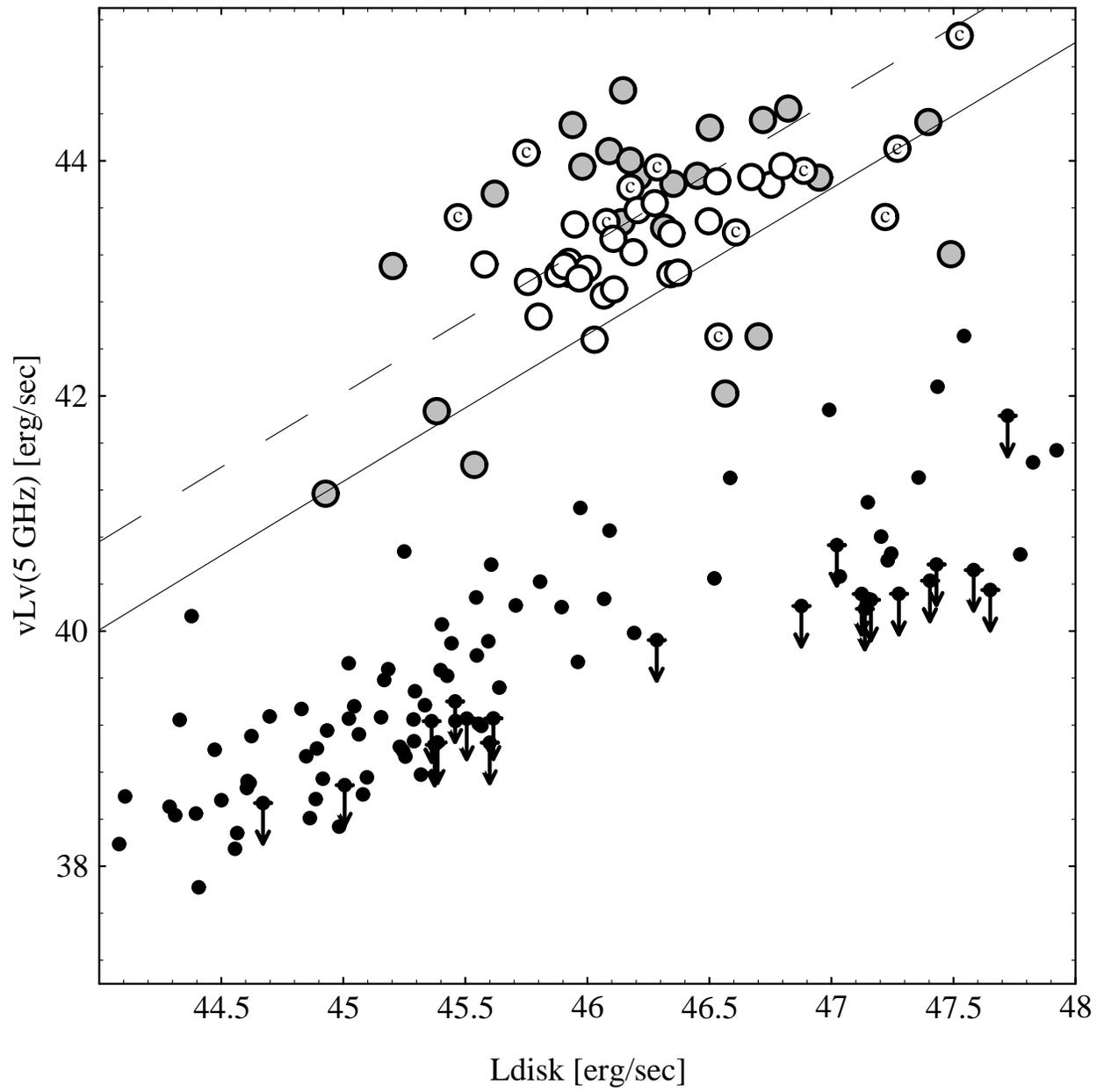
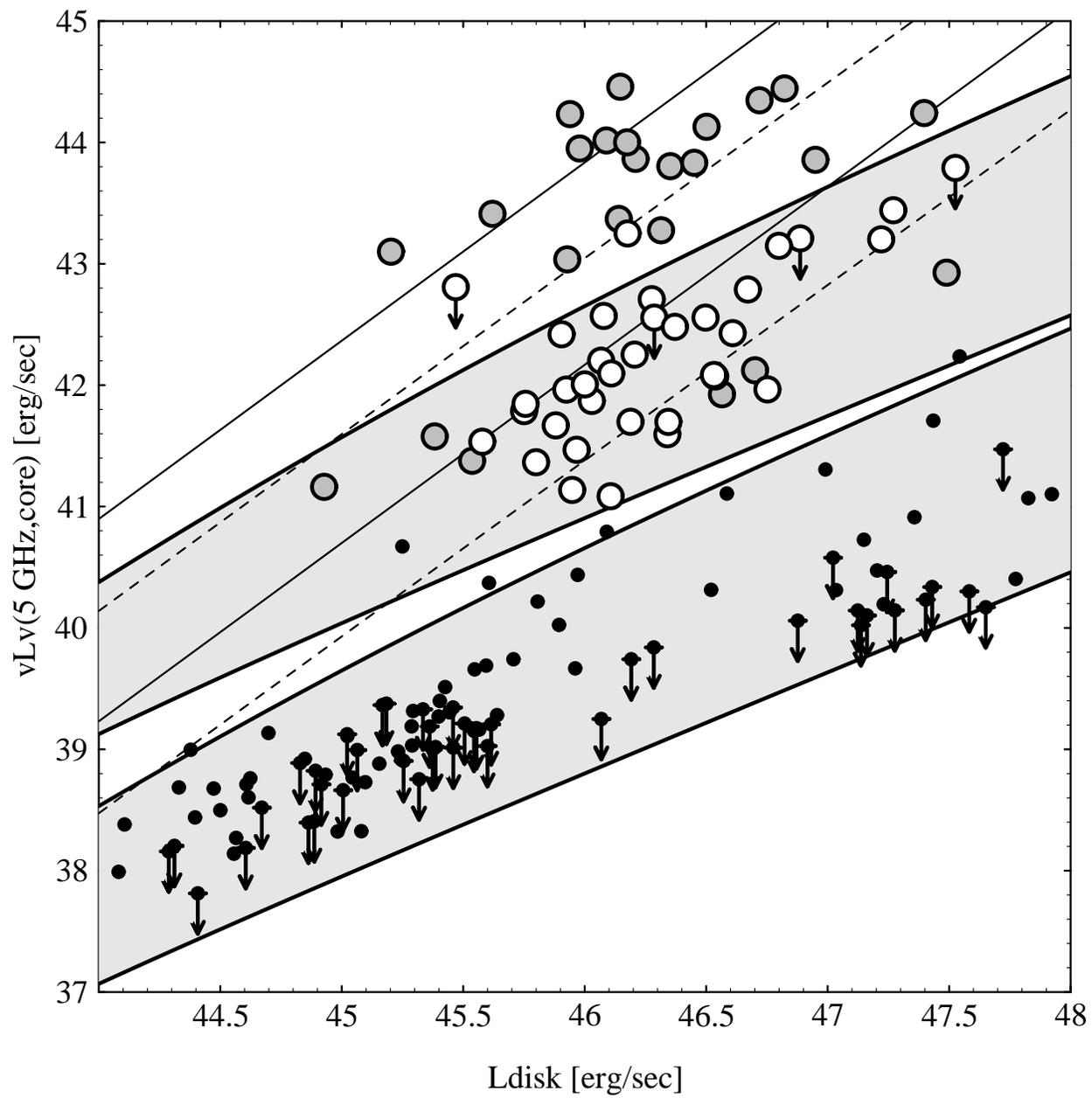
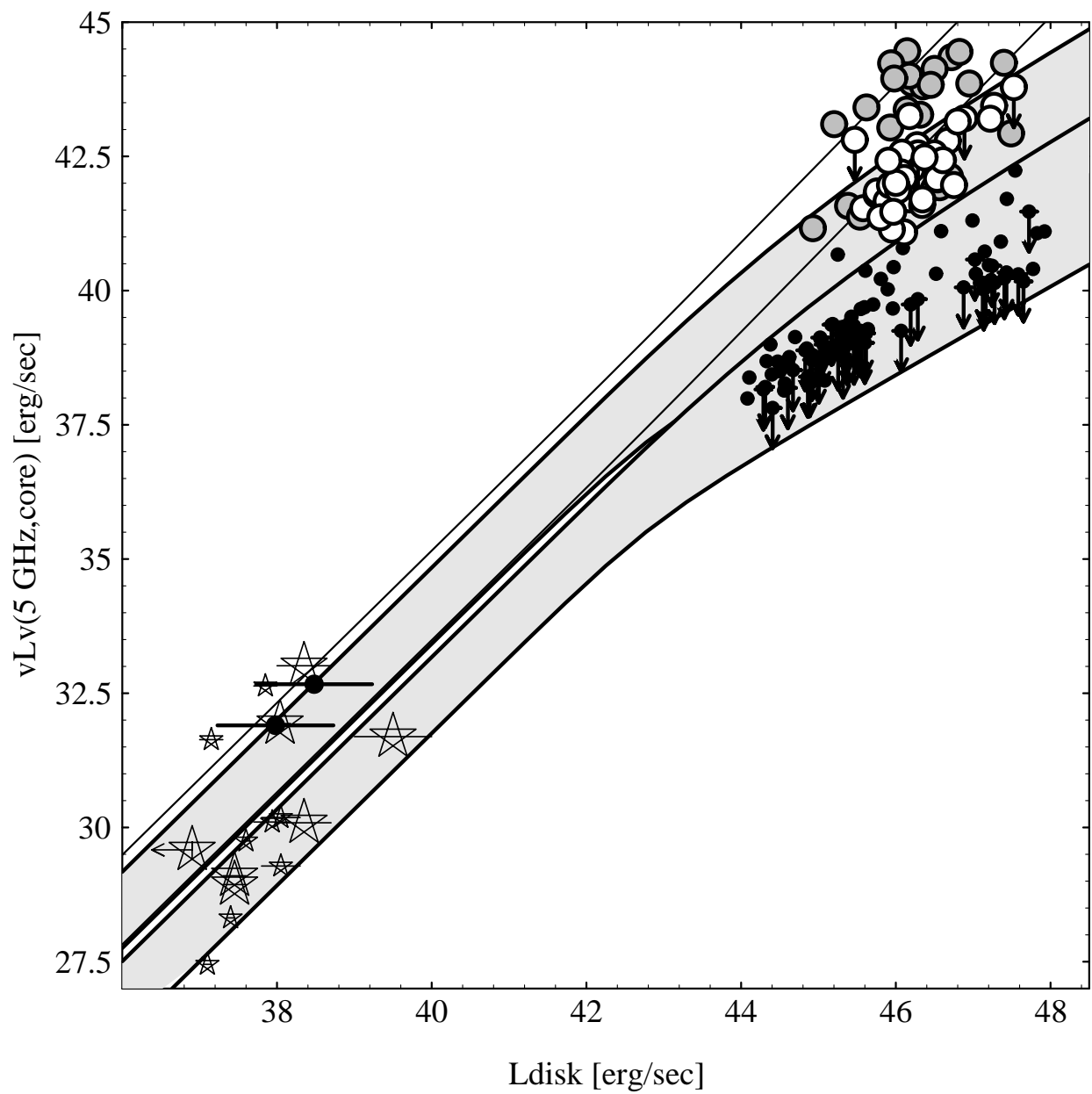
